\documentclass[12pt]{iopart}

\usepackage{iopams}  
\usepackage{epsfig}

\usepackage{amssymb,amsthm,amsmath}

\begin{document}

\title[Magnetic dipole transition in proton-deuteron radiative capture]{Magnetic dipole transition in proton-deuteron radiative capture at BBN energies within potential model}

\author{Nguyen Le Anh$^1$, Dao Nhut Anh$^1$, Do Huy Tho$^1$ and Nguyen Huu Nha$^2$}

\address{$^1$ Department of Physics, Ho Chi Minh City University of Education, 280 An Duong Vuong, District 5, Ho Chi Minh City, Vietnam}
\address{$^2$ Department of Theoretical Physics, Faculty of Physics and Engineering Physics, University of Science, Ho Chi Minh City, Vietnam}

\ead{anhnl@hcmue.edu.vn}

\begin{abstract}
The $pd$ radiative capture reaction plays a vital role in Big Bang nucleosynthesis and stellar proton-proton chain. The study of the low-energy reaction is challenging in both experiments and theories. Using the framework of potential model, we analyze $pd$ radiative capture below 1 MeV for both electric dipole ($E1$) and magnetic dipole ($M1$) transitions. The obtained astrophysical $S$ factors agree well with recent results, especially at energies relevant to sensitive deuterium abundance. The calculated reaction rate shows good agreement, with less than a 5\% difference compared to recent works.  The extrapolated value for $S(0)$ including both transitions is determined to be $0.211 \pm 0.016$ eV b. A comparison with experimental data using the $\chi^2$ test reveals the sensitivity of the $M1$ cross section at low energies to the scattering potential depth. 
\end{abstract}


\submitto{\PS}

\maketitle


\section{Introduction}
\label{introduction}

The primordial or Big Bang nucleosynthesis (BBN) is the production of the first nuclei of helium, lithium, and other light elements after the Big Bang. The formation and breakdown of deuterium involve a set of reactions, including the established $p$($n,\gamma$)$d$ pathway for deuterium production and the processes of deuterium reduction through $d$($d,n$)$^3$He, $d$($d,p$)$^3$H, and $d$($p,\gamma$)$^3$He ($pd$ radiative capture) reactions. The abundance of deuterium makes it a valuable indicator for cosmological parameters because it is highly responsive to the primordial baryon density \cite{cooke2018}. Furthermore, it is influenced by the number of neutrino species present in the early Universe \cite{xu2023}. In stars, the $pd$ radiative capture reaction is one of the key steps in the proton-proton chain, which converts hydrogen into helium and releases energy \cite{arcones2023}. The $pd$ radiative capture reaction in the BBN energy range has been extensively studied, especially in the last years, both theoretically and experimentally \cite{adelberger2011,cavanna2023}.

Due to the presence of the Coulomb barrier, the cross sections for the $pd$ radiative capture reaction at low energies are generally limited, making them challenging to accurately determine in experimental measurements. Ongoing and prospective laboratory experiments are actively exploring nuclear reaction physics, crucial components for input into BBN calculations. The $pd$ radiative capture reaction was measured at low energy in several accelerator experiments \cite{bystritsky2008,casella2002,griffiths1963,bailey1970,schmid1995,ma1997,weller1999,bystritsky2015,tisma2019,turkat2021,mossa2020}. In addition, high-energy-density plasmas provide an alternative technique for obtaining cross sections \cite{zylstra2020}. It is worth highlighting that this reaction is of great significance in astrophysics and is observed not only in the Sun but also in the majority of main-sequence stars throughout the universe. The greatest sensitivity of the primordial deuterium abundance to the $pd$ reaction cross section was particularly notable around 80 keV reported in Ref.~\cite{mossa2020}. Consequently, the experimental data observed below 1 MeV hold a pivotal role in understanding stellar processes in the main sequence and BBN.

In the early 2000s, the Laboratory for Underground Nuclear Astrophysics (LUNA) successfully conducted experiments to obtain important data on the astrophysical $S$ factor within the low-energy range \cite{casella2002}.  At higher energies, data sets were compiled from measurements in Refs.~\cite{griffiths1963,bailey1970,schmid1995,ma1997,bystritsky2015,tisma2019,turkat2021,mossa2020}. It is worth noting that these measurements retained considerable uncertainties, which considerably affected the comparison between predicted and observed primordial abundance. 
A new experimental campaign was relaunched at LUNA in 2016 \cite{cavanna2023}. 
The latest experiments conducted by LUNA have effectively decreased the uncertainty in the reaction rate to a level as low as 3\% \cite{mossa2020}. Several works, such as \cite{pisanti2021,moscoso2021}, discussed the astrophysical implications based on the updated $pd$ radiative capture rate.

Various theoretical models are extensively employed for extrapolating available data to extremely low energies, including $R$-matrix analysis, microscopic approaches, and potential models. The $R$-matrix analysis was compiled in Ref.~\cite{descouvemont2004}. The \textit{ab initio} methods recognized as state-of-the-art techniques for the $pd$ radiative capture reaction were reviewed in Ref.~\cite{adelberger2011}. The $pd$ scattering problem is an excellent testing ground for nuclear interactions, providing essential insights into the dynamics of few-body systems within the simplest nucleon-nucleus context. Advanced theoretical techniques with the help of Faddeev treatment or using hyperspherical harmonics (HH) expansions have been developed to tackle this challenge \cite{friar1991,kievsky1994,kievsky1995,viviani1996,golak2000,Sadeghi2013,vanasse2014,viviani2000,marcucci2005}. The nonlocality has been proposed to provide a better description for the three-nucleon bound states $^3$He and $pd$ scattering calculations \cite{doleschall2004,doleschall2005}. The recent theoretical work showed a disagreement at the level of 20–30\% with a widely used $S$-factor best fit to experimental datasets \cite{marcucci2016,marcucci2016erratum}. 

Besides the microscopic approach toward the solution of the general problem of nuclear forces, the $pd$ radiative capture can be addressed through a potential model that effectively reproduces experimental observations below the three-body breakup threshold \cite{huang2010,xu2013,ghasemi2018,dubovichenko2009,dubovichenko2020}. Within the potential model, the radiative capture process can be simply considered as an electromagnetic transition from the single-particle (s.p.) scattering state to the s.p. bound state \cite{huang2010}. The $pd$ radiative capture reaction in this work is examined using a phenomenological approach reported in Refs.~\cite{huang2010,xu2013,ghasemi2018} to obtain astrophysical $S$ factors, reaction rates, and the extrapolated value of $S(0)$. The process of $pd$ radiative capture is complicated, involving both electric dipole ($E1$) and magnetic dipole ($M1$) transitions, with the latter making a significant contribution. Thus, our main focus is to evaluate the contribution of the $M1$ transition within the potential model. The role of electromagnetic dipole and high-order transitions has been extensively studied with the pair-correlated HH method for $pd$ radiative capture in the literature \cite{viviani2000}. The comparison between the experimental data and the theoretical \textit{ab initio} calculation for the three-body problem including the meson exchange currents (MEC) effects shows the sensitivity of $M1$ transitions \cite{ma1997,viviani1996,rice1997}. In the present work, the sensitivity of the $M1$ transition is also pointed out within the potential model using the well-depth method. We show that describing nuclear three-body systems as a two-body system could provide not only a simple and effective way to describe experimental data but also key inputs of nuclear astrophysics \cite{huang2010,xu2013}. Also, the nuclear spectroscopic information in $^3$He and $pd$ scattering observables are revealed.

The structure of this paper is as follows: In the following section, we will present the fundamental formula for a radiative capture reaction. The detailed formulation of the $E1$ and $M1$ transitions within the potential model can be found in Ref.~\cite{anh2022PRC1061}. The obtained astrophysical $S$ factors, reaction rates, and elastic scattering observables are discussed in the result section.

\section{Formalism for radiative capture reaction within potential model}

\subsection{Reaction rate and astrophysical $S$ factor}
The astrophysical reaction rates per particle pair at a certain temperature $T$ can be calculated with
\begin{equation} \label{eq:rate}
    \langle \sigma v \rangle  = \sqrt{\dfrac{8}{\pi \mu}}\dfrac{1}{(k_BT)^{3/2}} \int_0^\infty g(E)S(E)  \,dE,
\end{equation}
where $k_B$ is the Boltzmann constant and $\mu$ is the reduced mass of the $pd$ system. The masses of proton and deuteron used in this work are $1.0073$~u and $2.0141$~u, respectively. The Gamow window function is expressed as
\begin{equation} \label{eq:GM}
    g(E)= \exp \left[-\dfrac{E}{k_BT}-2\pi\eta(E)  \right],
\end{equation}
where the Sommerfeld parameter is $\eta(E) = e^2/(\hbar v)$ with $v$ being the relative velocity between the proton and the deuteron.  As the energy approaches zero, the cross sections $\sigma(E)$ exhibit a significant decrease. It is therefore customary to introduce the energy-dependent astrophysical $S$ factor defined as
\begin{equation} \label{eq:S_def}
    S(E) = E e^{2\pi\eta} \sigma(E).
\end{equation}
When $S(E)$ remains constant, which occurs in the absence of resonances, the integrand in Eq. \eqref{eq:rate} reaches its peak efficiency at the most effective energy.

To evaluate the agreement between the calculated $S(E)$ and experimental data, the $\chi^2$ statistic is defined as
\begin{equation}
    \chi^2 = \sum_{i=1}^N \left(\dfrac{S_i^{\text{cal}}-S_i^{\text{exp}}}{\Delta S_i^{\text{exp}}}\right)^2,
\end{equation}
where $S_i^{\text{cal}}$ and $S_i^{\text{exp}}$ are the calculated and experimental astrophysical $S$ factors for a set of $N$ selected data points from experiments, respectively. These selected experimental data points below 1 MeV are from Refs.~\cite{casella2002,griffiths1963,bailey1970,schmid1995,ma1997,bystritsky2015,tisma2019,turkat2021,mossa2020}, and each data point is associated with an uncertainty denoted as $\Delta S_i^{\text{exp}}$.

In the potential model, the internal structure of the interacting nuclei is essentially neglected. The intrinsic spins of the target ($I$) and incident proton ($s=1/2$) are thus kept unchanged. The system is assumed as the core (target) capturing a proton into the s.p. state.  The initial (scattering) state is denoted as $| [I \otimes (\ell_i \otimes s){j_i}]{J_i}\rangle$, while the final (bound) state is represented as $| [I \otimes (\ell_f\otimes s){j_f}]{J_f} \rangle$.  The total relative angular momentum of the system is $\Vec{j} = \Vec{\ell} + \Vec{s}$ with $\Vec{\ell}$ being the relative orbital angular momentum. The channel spin is a result of coupling $\Vec{J} = \Vec{I} + \Vec{j}$. The radiative capture cross section for the electromagnetic dipole ($\Omega 1$ with $\Omega \equiv E$ or $\Omega \equiv M$) transitions to a bound state is now written as 
\begin{equation}\label{eq:sigmaE}
        \sigma_{n \ell_f j_f J_f} (E) = \dfrac{4}{3}\dfrac{1}{\hbar v} \left(\dfrac{4\pi}{3} k^3_\gamma \right) \dfrac{1}{(2s+1)(2I+1)}
        \times \sum_{\Omega,\ell_i j_i J_i} |M_{\Omega 1}|^2 ,
\end{equation}
where the $\gamma$-ray wave number is defined as
\begin{equation}
    k_\gamma = \dfrac{E-E_{n\ell_f j_f}}{\hbar c},
\end{equation}
as a function of the proton energy $E$. The binding energy $E_{n\ell_f j_f}$ is determined within the potential model.

To determine the capture cross sections as described in Eq.~\eqref{eq:sigmaE}, it is necessary to compute the matrix elements of the dipole operators. The reduced matrix elements of the $\Omega 1$ transition are written as
\begin{equation}\label{Mstart}
    M_{\Omega 1} 
    = \langle [I \otimes (\ell_f \otimes s)_{j_f}]_{J_f} || \mathcal{O}_{\Omega 1} || [I \otimes (\ell_i \otimes s)_{j_i}]_{J_i}\rangle.
\end{equation}
The dipole operators of $E1$ and $M1$ take the forms \cite{anh2022PRC1061}
\begin{align}
    \mathcal{O}_{E1} &= C_e r Y_1,\\
    \mathcal{O}_{M1} &= \sqrt{\dfrac{3}{4\pi}} \left(C_m \hat{\ell} + 2\mu_p \hat{s} + 2\mu_d \hat{I}  \right),
\end{align}
where $Y_1$ are spherical harmonics, and the hat notations are the projections of the orbital angular momentum and spins. The values of $C_e$ and $C_m$ are the effective charge and effective magnetic moment, respectively. The magnetic moments that are used in this work are $\mu_p =2.79285\mu_N$ for proton and $\mu_d = 0.85744\mu_N$ for deuteron \cite{mohr2000}, where $\mu_N$ is the nuclear magneton.

The $M_{\Omega 1}$ in Eq.~\eqref{Mstart} can be simplified by the calculation of the s.p. reduced matrix element which can be decomposed into components
\begin{equation} \label{eq:spME1}
    M_{\Omega 1}^{({\text{s.p.}})} = A \cdot I \cdot \sqrt{S_{\rm F}},
\end{equation}
where $S_{\rm F}$ is introduced to account for the fractional parentage coefficient that the system can be described as the presumption. The formula of $E1$ and $M1$ transitions relating to angular-spin coefficients $A$ can be found in Ref.~\cite{anh2022PRC1061}. The most important ingredient of Eq.~\eqref{eq:spME1} is the radial overlap integrals $I$ of two states.
\begin{align}\label{eq:I1}
    I_{E1} &= \int \phi_{n \ell_f j_f}(r) \chi_{\ell_i j_i}(E,r)r \,dr, \\
    I_{M1} &= \int \phi_{n \ell_f j_f}(r) \chi_{\ell_i j_i}(E,r) \,dr,
\end{align}
where $\chi_{\ell_i j_i}$ and $\phi_{n \ell_f j_f}$ are s.p. wave functions of scattering and bound states, respectively.

\subsection{Phenomenological potential model}
The bound and scattering wave functions are described as eigenfunctions of the usual radial Schr\"odinger equations for each s.p. wave function
\begin{align}
    \left[ -\dfrac{\hbar^2}{2\mu} \left(\dfrac{d^2}{dr^2}-\dfrac{\ell_f(\ell_f+1)}{r^2} \right) + V_{n\ell_f j_f}(r) \right] \phi_{n\ell_f j_f}(r) &= E_{n\ell_f j_f} \phi_{n \ell_f j_f}(r), \label{eq:eqbound} \\
    \left[ -\dfrac{\hbar^2}{2\mu} \left(\dfrac{d^2}{dr^2}-\dfrac{\ell_i(\ell_i+1)}{r^2} \right) + V_{\ell_i j_i}(r) \right] \chi_{\ell_i j_i}(E,r)
    &= E \chi_{\ell_i j_i}(E,r) .\label{eq:eqscatt}
\end{align}
In both Eqs. \eqref{eq:eqbound} and \eqref{eq:eqscatt}, the local potentials $V_{n\ell_f j_f}(r)$ and $V_{\ell_i j_i}(r)$ have identical forms and include contributions from nuclear central, spin-orbit coupling, and Coulomb terms
\begin{equation}
    V(r) = V_{\text{cent.}}(r) + V_{\text{s.o.}}(r)(\vec{\ell}\cdot\vec{\sigma}) + V_{\text{Coul.}}(r). \label{eq:Vbound} 
\end{equation}
The nuclear terms are of the Woods-Saxon (WS) form
\begin{align}
    V_{\text{cent.}}(r) &=  V_0 f_{\text{WS}} \label{eq:Vcentpheno} ,\\
    V_{\text{s.o.}} (r) &= \left( \dfrac{\hbar}{m_\pi c} \right)^2 \dfrac{V_{{\text{S}}}}{r} \dfrac{d}{dr} f_{\text{WS}} \label{eq:Vsopheno} ,\\
    f_{\text{WS}}(r) &=  \dfrac{1}{1+e^{(r-R_{0})/a_{0}}},
\end{align}
where $R_0$ and $a_0$ are the radius and diffuseness parameters of the WS potentials, respectively. The value of $m_\pi$ is the pion rest mass. The parameters $V_0$ and $V_{\text{S}}$ are the strengths (depths) of the nuclear central and spin-orbit potentials in MeV, respectively. The repulsive part of the central potential is omitted. While $V_0$ is a negative value, the $V_{\text{S}}$ is chosen with a positive value for consistency with the shell model of the nucleus. The Coulomb potential is of a uniformly charged sphere
\begin{align} \label{eq:Vcoulpheno}
    V_{\rm Coul.}(r) = \begin{cases}
        \dfrac{e^2}{2R_{{\text{C}}}}\left( 3- \dfrac{r^2}{R_{{\text{C}}}^2}\right), & r < R_{{\text{C}}} \\
        e^2/r, &r \geq R_{{\text{C}}}
    \end{cases} ,
\end{align}
where $R_{\text{C}}$ is the Coulomb radius.

The parameters $V_0$ and $V_{\text{S}}$ are fine-tuned to reproduce the ground-state energy $E_{n\ell_f j_f}$ of the deuteron in Eq.~\eqref{eq:eqbound}. The wave function of the bound state $\phi_{n\ell_f j_f} (r)$ becomes negligible at large distances, and its norm is determined by
\begin{equation}
    \langle \phi_{n\ell_f j_f}|\phi_{n\ell_f j_f}\rangle = 1.
\end{equation}

The potential parameters for the scattering states are similar to those for the bound states. The scattering states $\chi_{\ell_i j_i}(E, r)$ with continuous energy $E$ satisfy boundary conditions at infinity replaced by
\begin{equation} \label{eq:matching}
    \chi_{\ell_i j_i}(E,r \to \infty) \to \cos \delta_{\ell_i j_i}(E) F_{\ell_i}(\eta,kr) + \sin\delta_{\ell_i j_i}(E) G_{\ell_i}(\eta,kr),
\end{equation}
where $F_{\ell_i}(\eta,kr)$ and $G_{\ell_i}(\eta,kr)$ are the regular and irregular Coulomb wave functions, respectively; and $\delta_{\ell_i j_i}(E)$ represents the nuclear phase shift for the partial wave $\ell_i j_i$. 

\section{Results and discussions} 
\subsection{Single-particle configuration and potential parameters}

\begin{figure}
    \centering
    \includegraphics[width=0.80\textwidth]{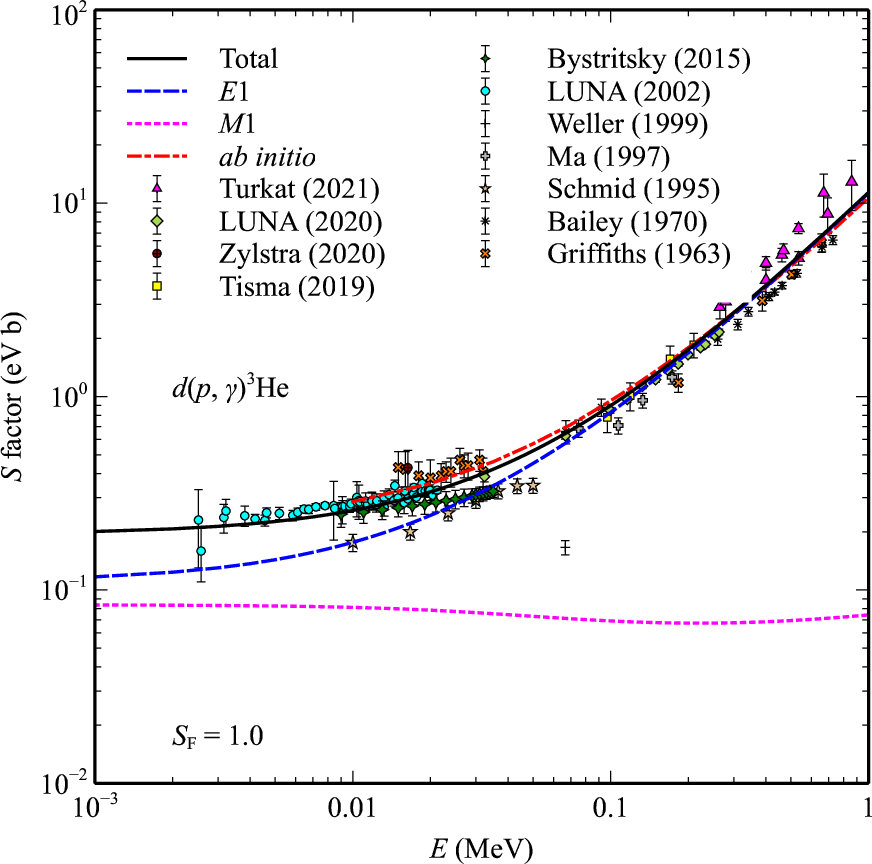}
    \caption{Astrophysical $S$ factor of $d$($p,\gamma$)$^{3}$He reaction with experimental data selected from Refs.~\cite{casella2002,griffiths1963,bailey1970,schmid1995,ma1997,weller1999,bystritsky2015,tisma2019,turkat2021,mossa2020,zylstra2020}. The solid line represents the total calculation with $E1$ (dashed line) and $M1$ (dotted line) transitions. The dash-dotted line shows the astrophysical $S$ factor using the theoretical \textit{ab initio} model in Ref.~\cite{marcucci2016}.}
    \label{fig:sfactor}
\end{figure}

There are no excited states in the deuteron, and there is no evidence supporting a low-lying excited state in $^3$He \cite{tilley1987}. We therefore consider the transitions to the ground state of $^3$He with $J_f = 1/2^+$. The ground state of $^3$He is modeled as a proton with spin $s = 1/2$ coupled to the deuterium core, which itself possesses an intrinsic spin $I = 1^+$. In the framework of the shell model, it is conventional to describe $^3$He within a model space characterized by $0\hbar\omega$, with approximately three nucleons in the $1s_{1/2}$ states. The correlation in the $^3$He ground state, resulting from the mixing of higher components, is estimated to be approximately below 10\% in shell-model calculation \cite{dortmans1998}. 

In this work, the form of WS potential is fitted based on the form of Skyrme Hartree-Fock potential \cite{colo2013}. The version of SLy5 interaction gives the values of $R_0=2.31$ fm and $a_0=0.37$ fm for the central potential. To simplify the model and reduce the number of parameters, we maintain constant values for the parameters $a_0=0.37$ fm, $R_0 =r_{\text{C}}=2.31$ fm, and $V_{\text{S}} = 5$ MeV. It is important to note that the diffuseness and radius parameters have minimal impact on the shape of cross section. Additionally, given the absence of resonances in the energy region below 1 MeV, adjustments to the strength of the spin-orbit interaction do not significantly affect the cross section. Consequently,  only adjustment of the depth $V_0$ is used to reproduce simultaneously the binding energy and the $pd$ scattering.

For the bound state, an additional proton is captured into the unoccupied $1s_{1/2}$ state ($\ell_f=0$). The binding energy of $1s_{1/2}$ proton is equal to the $Q$ value of the reaction, which is $E_{1s_{1/2}} =-5.49$ MeV. To reproduce this binding energy, the value of $V_0^b = -32$ MeV is adopted. The $s$-wave asymptotic normalization constant (ANC) in $^{3}$He is determined to be $1.79$ fm$^{-1/2}$ in this work, yielding a squared value of $3.2$ fm$^{-1}$. This value is slightly below measurements by less than 10\% but falls within the uncertainty of measurements. In particular, the empirical squared ANCs reported are $3.4\pm 0.2$ fm$^{-1}$ in Ref.~\cite{plattner1977} and $3.5\pm 0.4$ fm$^{-1}$ in Ref.~\cite{bornand1978}. Additionally, the ANC computed for the Reid soft-core potential is found to be 1.765 fm$^{-1/2}$ \cite{sasakawa1980} which is close to the value obtained from our potential model. 

The $E1$ transition is predominantly caused by incoming $p$ waves ($\ell_i=1$). The channel spins corresponding to possible captured $p$ waves (both $p_{1/2}$ and $p_{3/2}$) are $J_i=1/2^-, 3/2^-$. Fig.~\ref{fig:sfactor} shows the calculated astrophysical $S$ factor of $pd$ radiative capture. For comparison, the dash-dotted line represents the astrophysical $S$ factor obtained through the \textit{ab initio} approach, extracted from Table I in Ref.~\cite{marcucci2016}. The dashed line presents our calculation with only the $E1$ transition. Notably, it well describes the data sets from Refs.~\cite{schmid1995,tisma2019,turkat2021,mossa2020} without modification. However, it is still lower than the LUNA database at very low energy \cite{casella2002}. The extrapolated value of $S_{E1}(0)$ with only $E1$ contribution is 0.16 eV b, using $V_0^s=V_0^b=-32$ MeV.

\subsection{Sensitivity of $M1$ transition}

\begin{figure}
    \centering
    \includegraphics[width=0.80\textwidth]{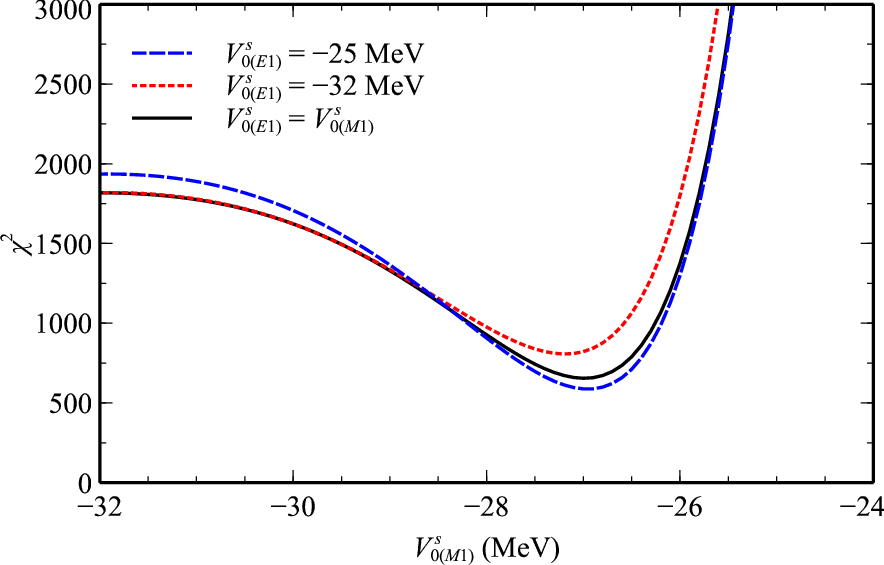}
    \caption{Results of the $\chi^2$ test examining the variation of the depth $V^s_{0(M1)}$ of the scattering potential for the $s$ state. The optimal depth is found to be $V^s_{0(M1)}=-27$ MeV, with values of $a_0=0.37$ fm and $R_0 = 2.31$ fm. The dashed line represents the calculation using $V_{0(E1)}^s = -25$ MeV for scattering states in the $E1$ transition. The dotted line illustrates the calculation using $V_{0(E1)}^s = -32$ MeV, consistent with the bound state. The solid line shows the calculation employing the same scattering potential depths for both $E1$ and $M1$ transitions.}
    \label{fig:chi2}
\end{figure}

The contribution of the $M1$ transition is considered, which enhances the magnitude of the $S$ factor at low energy. The scattering $s$ wave ($\ell_i=0$) causes the $M1$ transition to the ground state. Fig.~\ref{fig:chi2} shows the optimal value $V_0$ when using $\chi^2$ test in comparison with experimental data from Refs.~\cite{casella2002,griffiths1963,bailey1970,schmid1995,ma1997,bystritsky2015,tisma2019,turkat2021,mossa2020} as displayed in Fig.~\ref{fig:sfactor}. It is evident that the minimum $\chi^2$ value occurs when the depth $V_0$ is around $-27$ MeV. It is reasonable when $V_0^s = -27$ MeV is applied to both scattering $s$ and $p$ waves. As the depth increases beyond this value, the $\chi^2$ value remains relatively stable, primarily because the $M1$ contribution is negligible. For instance, when the same potentials for bound and scattering states ($V^s_0= V^b_0=-32$ MeV) are adopted, the $S$ factor for the $M1$ transition is lower by approximately 9 orders of magnitude compared to the primary contribution from the $E1$ transition. In contrast, decreasing the depth leads to an overestimation of the $M1$ strength, highlighting the high sensitivity of the $M1$ transition strength to variations in potential depth. 

\subsection{Analysis of elastic cross section and polarization}

\begin{figure}
    \centering
    \includegraphics[width=0.80\textwidth]{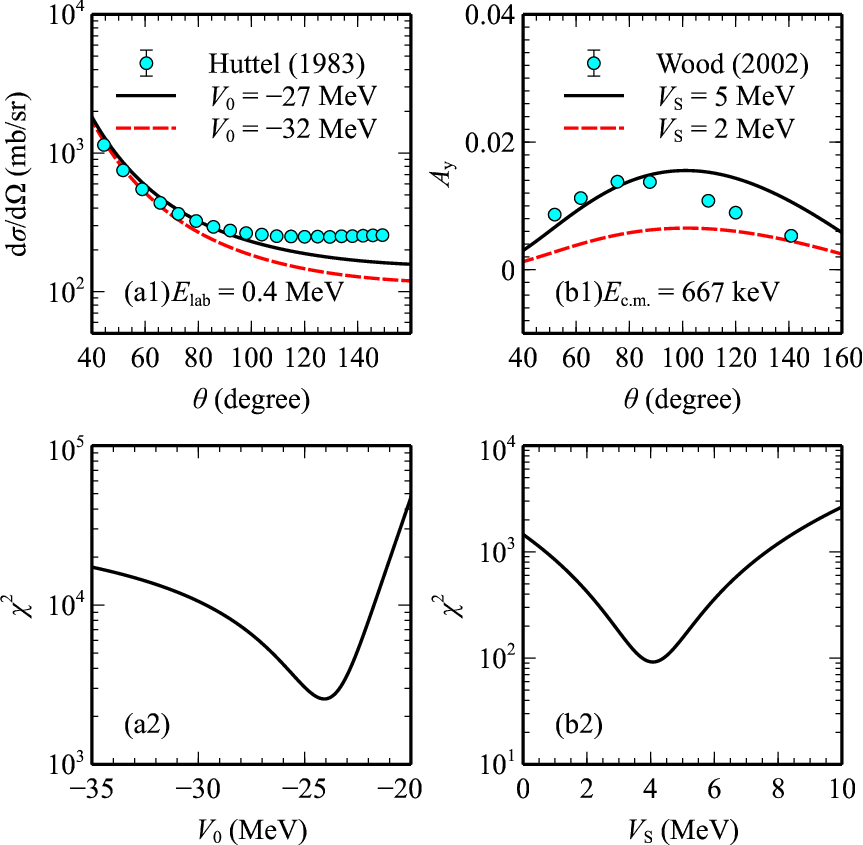}
    \caption{(a1) Angular distributions of differential cross sections for $pd$ elastic scattering at $E_{\text{lab}}=0.4$ MeV, with experimental data sourced from Ref.~\cite{huttel1983}. (b1) Angular distributions of the analyzing power calculated using $V_{0}=-27$ MeV with $V_{\text{S}}=2$ MeV (dashed line) and $V_{\text{S}}=5$ MeV (solid line). Data for the analyzing powers at $E_{\text{c.m.}}=667$ keV are taken from Ref.~\cite{wood2002}. (a2, b2) The $\chi^2$ test examining the variations of the depths of central potential $V_0$ and spin-orbit potential $V_{\rm S}$, respectively.}.
    \label{fig:ay_cs}
\end{figure}

Validating the scattering potentials requires the calculation of cross sections and polarizations at low energy. The differential cross section for $pd$ elastic scattering at $E_{\text{lab}}=0.4$ MeV is depicted in Fig.~\ref{fig:ay_cs}(a1), using measurements of angular distributions from Ref.~\cite{huttel1983}. The $\chi^2$ value decreases by a factor of 2.7 when using $V_0 = -27$ MeV compared to $V_0 = -32$ MeV. The optimal depth of the central potential is determined to be approximately $-25$ MeV, as shown in Fig.~\ref{fig:ay_cs}(a2). Notably, an improvement in the cross section is observed when employing the central scattering potential $V_0=-27$ MeV for large scattering angles, with the inclusion of a spin-orbit potential $V_{\text{S}}=5$ MeV. Although this inclusion does not have a significant impact on the description of differential cross section, it does play a role in investigating polarization.

In Fig. \ref{fig:ay_cs}(b1), the calculated analyzing powers are presented with varying strengths of the spin-orbit potential. The strength of $V_{\text{S}}=5$ MeV aligns well with data points at 667 keV from Ref.~\cite{wood2002}. In contrast, a spin-orbit strength of $V_{\text{S}}=2$ MeV fails to replicate in this case. The $\chi^2$ value for $V_{\rm S} = 5$ MeV is reduced by half compared to $V_{\rm S} = 2$ MeV. Fig.~\ref{fig:ay_cs}(b2) reveals that the optimal depth of the spin-orbit potential is approximately $4$ MeV. The description of $pd$ scattering observables, especially the angular distributions of the analyzing powers, is enhanced when including MEC effects, as reported in Ref.~\cite{ma1997}. It is important to note that the optical potentials including central and spin-orbit potentials in this study are considered only real and energy-independent at low energy. Therefore, the examination for higher energies is not within the scope of this work.

\subsection{Spectroscopic factor and the best-fit value of $S(0)$}

\begin{figure}
    \centering
    \includegraphics[width=0.80\textwidth]{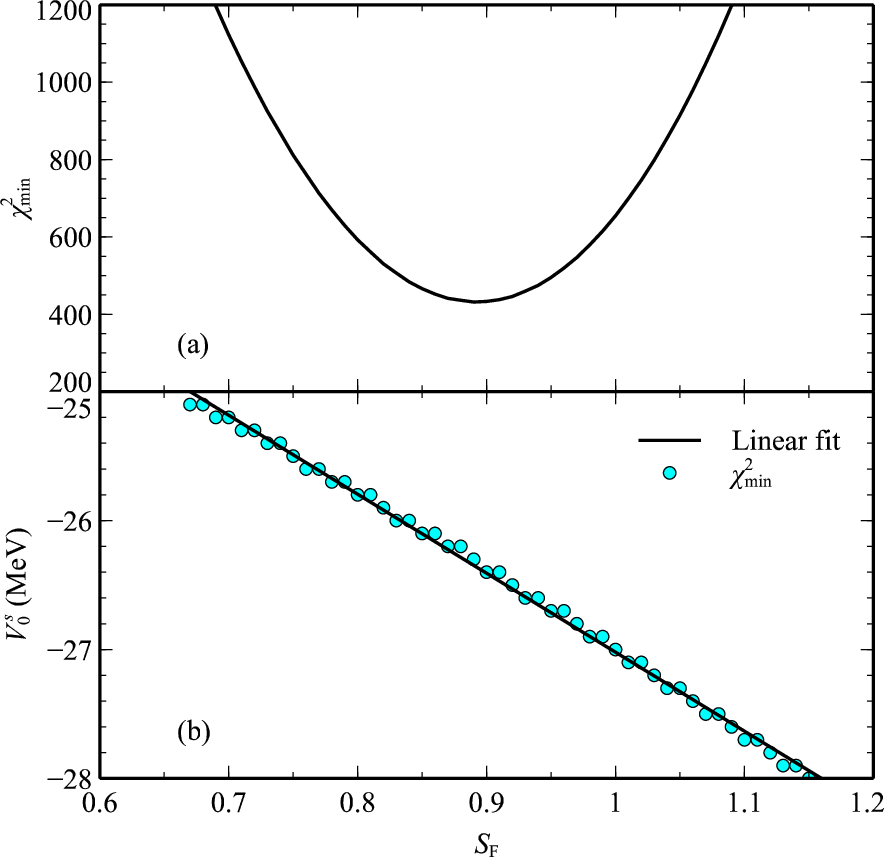}
    \caption{(a) The different $\chi^2_{\min}$ values obtained when changing $S_{\rm F}$ and $V_0^s$. The best value for the proton spectroscopic factor in $^3$He is $S_{\rm F}=0.9$ corresponding to the $V_0^s = -26.3$ MeV, which gives the lowest value of $\chi^2$. (b) The linear correlation between $S_{\rm F}$ and $V_0^{s}$ at the minimum $\chi^2$ expressed by $V_{0}^s = -21.9 - 6.1 S_{\rm F}$ in MeV.}
    \label{fig:chi2_min}
\end{figure}

The spectroscopic factor $S_{\rm F}$ for the bound state is determined by finding the values of $\chi^2_{\min}$.  In Fig.~\ref{fig:chi2_min}(a), the variations in $\chi^2_{\min}$ are depicted concerning changes in $S_{\rm F}$ and $V_0^s$. The best-fit value for $S_{\rm F}$ is identified as $0.9$, corresponding to the use of $V_0^s = -26.3$ MeV for both transitions. At the minimum of $\chi^2$, the relationship between $S_{\rm F}$ and $V_0^{s}$ follows a linear pattern given by $V_{0}^s = -21.9 - 6.1 S_{\rm F}$ in MeV, as depicted in Fig.~\ref{fig:chi2_min}(b). The revised $S_{E1}(0)$ is now $0.098$ eV b. In comparison, Ref.~\cite{descouvemont2004}, utilizing $R$-matrix analysis, reported $S_{E1}(0) = 0.089 \pm 0.004$ eV b, which closely aligns with our calculated value.

\begin{figure}
    \centering
    \includegraphics[width=0.80\textwidth]{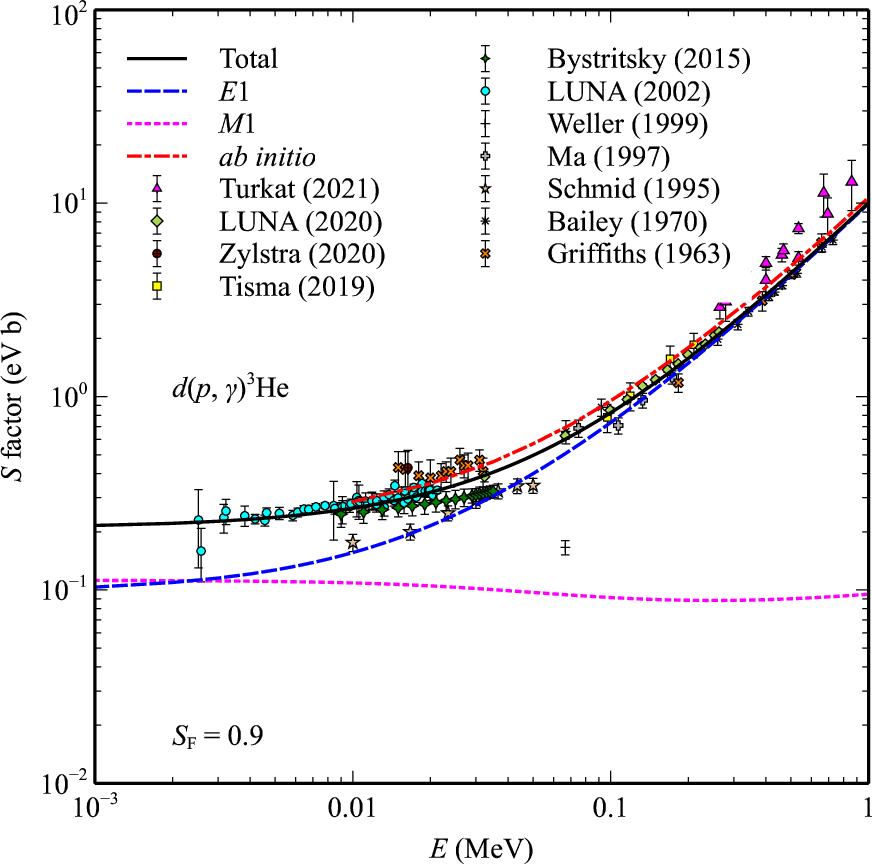}
    \caption{Same as Fig.~\ref{fig:sfactor} but for $S_{\rm F}=0.9$ and $V_0^s = -26.3$ MeV.}
    \label{fig:sfactor_SF089}
\end{figure}

\begin{table}
    \centering
    \caption{The values of $S(0)$ for different experimental and theoretical works.}
    \label{tab:S0}
    \begin{tabular}{ll} \\
    \hline \hline
        References & $S(0)$ (eV b) \\ \hline
        Schmid \textit{et al.} (1996) \cite{schmid1996} & $0.165 \pm 0.014$ \\
        Viviani \textit{et al.} (1996) \cite{viviani1996} & $0.185$ \\
        Viviani \textit{et al.} (2000) \cite{viviani2000} & $0.219$ \\
        Casella \textit{et al.} (2002) \cite{casella2002} & $0.216 \pm 0.010$ \\
        Descouvemont \textit{et al.} (2004) \cite{descouvemont2004} &  $0.223 \pm 0.010$  \\
        Marcucci \textit{et al.} (2005) \cite{marcucci2005} & $0.219$ \\
        Xu \textit{et al.} (2013) \cite{xu2013} & $0.21\pm 0.04$ \\
        Iliadis \textit{et al.} (2016) \cite{iliadis2016} & $0.2156^{+0.0082}_{-0.0077}$  \\
        Sadeghi \textit{et al.} (2013) \cite{Sadeghi2013} & 0.243 \\
        Turkat \textit{et al.} (2021) \cite{turkat2021} & $0.219 \pm 0.004$ \\
        Moscoso \textit{et al.} (2021) \cite{moscoso2021} & $0.219 \pm 0.001$ \\
        This work & $0.211 \pm 0.016$ \\
        \hline \hline
    \end{tabular}
\end{table}

In Fig.~\ref{fig:sfactor_SF089}, the inclusion of $M1$ transition leads to an extrapolated value of $S(0)$ of $0.211 \pm 0.016$ eV b, using $V_0^s=-26.3$ MeV and $S_{\rm F}=0.9$. The uncertainty in our computed $S$ factor arises from the variance between calculations utilizing $S_{\rm F}=0.9$ and $S_{\rm F}=1.0$. The contribution of $M1$ in our calculation at zero energy is determined to be $S_{M1}(0)= 0.113$ eV which is in excellent agreement with the experimental determination in Refs.~\cite{griffiths1963} and \cite{schmid1995} reported as $0.12 \pm 0.03$ eV b and $0.109 \pm 0.010$ eV b, respectively. The enhancement of $M1$ transition at very low energy is due to two-body current contributions \cite{schmid1995}. Table \ref{tab:S0} presents our results of the total $S(0)$ compared with the other works. The experimental references have reported values such as $S(0)=0.166\pm 0.014$ eV b \cite{schmid1995}, $0.216\pm 0.010$ eV b \cite{casella2002}, and $0.219 \pm 0.004$ eV b \cite{turkat2021}. The calculation of $S(0)$ with the inclusion of $M1$ contribution is $4\%$ lower than recent experimental values in Refs.~\cite{casella2002,turkat2021} and slightly smaller than the \textit{ab initio} value of 0.219 eV b \cite{viviani2000,marcucci2005}. Using $R$-matrix analysis, $S(0)$ is found to be $0.089 \pm 0.004$ eV b ($E1$) and $0.134 \pm 0.006$ eV b ($M1$) \cite{descouvemont2004}.  Our calculation shows a good agreement with the value reported in the NACRE II compilation~\cite{xu2013}.  It is worth mentioning that the fit of potential depth for both $E1$ and $M1$ transitions is significantly influenced by very low-energy data points obtained from measurements in Ref.~\cite{casella2002}. Specifically, a data point at 4.05 keV (see Table I in Ref.~\cite{casella2002}) stands out significantly lower compared to the trend of other data points as shown in Figs.~\ref{fig:sfactor} and \ref{fig:sfactor_SF089}. 

Around the point where it is emphasized that the primordial deuterium abundance is most sensitive ($E=80$ keV) according to Ref.~\cite{mossa2020}, our calculated $S$ factors are in good agreement with the latest measured values. Particularly, the total $S$ factors at $E=66.7$ keV and $E=99.5$ keV in the present work give $0.602 \pm 0.042$ eV b and $0.821 \pm 0.073$ eV b while the experimental values are $0.627 \pm 0.025$ eV b and $0.850\pm 0.029$ eV b \cite{mossa2020}, respectively. The prediction of $S$ factor at 91 keV reported in Ref.~\cite{moscoso2021} is $0.799 \pm 0.018$ eV b which is slightly higher than our value of $0.763 \pm 0.065$ eV b. In this range of energy, the contribution of the $M1$ transition in our approach is approximately 11\% to 16\%.

At very low energy, the total $S$ factor (in eV b) can be approximated as a polynomial function of energy (in MeV). For energies below 40 keV, Ref.~\cite{griffiths1963} reported the linear function $S(E) = (0.25 \pm 0.04) + 7.9E$. Additionally, a cubic function  represented as $S(E)= 0.2121 + 5.973E + 5.449 E^2 -1.656E^3$ was reported for energies below 2 MeV in Ref.~\cite{mossa2020}. Based on the statistical model with data from 11 experiments, Ref.~\cite{moscoso2021} gave $S(E)= 0.219^{+0.01}_{-0.01} + 5.8^{+0.24}_{-0.24}E + 6.34^{+0.88}_{-0.82} E^2 -2.2^{+0.52}_{-0.52}E^3$. In the present work, the total $S$ factor calculated for energies below 100 keV is approximated by
\begin{equation}
    S(E) \approx 0.211 + 5.25E + 9.47E^2 -5.54E^3,
\end{equation}
where $S(E)$ and $E$ are in eV b and MeV, respectively. The slope of $S(E)$ from our calculation is slightly lower than those of Refs.~\cite{mossa2020,moscoso2021}. The approximated $S$ factors for $E1$ and $M1$ transitions using $V_0^s=-26.3$ MeV and $S_{\rm F}=0.9$ are given by
\begin{align}
    S_{E1}(E) \approx 0.098 + 5.72E + 5.94E^2 + 4.59E^3,\\
    S_{M1}(E) \approx 0.113 - 0.46E + 3.54E^2 - 10.1E^3.
\end{align}
Notably, the slope of $S_{M1}(E)$ is negative, indicating that $S_{M1}(E)$ is influenced by a sub-threshold $s$-state resonance. In contrast, the positive slope of $S_{E1}(E)$ suggests $p$-state resonances above 1 MeV.

\subsection{Reaction rate}

\begin{figure}
    \centering
    \includegraphics[width=0.80\textwidth]{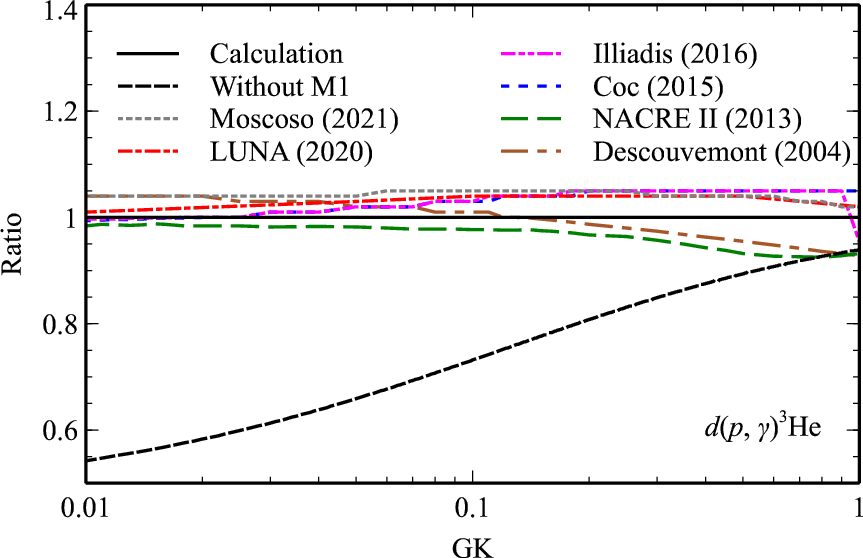}
    \caption{The ratio of reaction rates below 1 GK obtained from Refs.~\cite{mossa2020,moscoso2021,descouvemont2004,xu2013,iliadis2016,coc2015} to the calculated rates in this work.}
    \label{fig:rate}
\end{figure}

Our calculation of Gamow window functions $g(E)$ given in Eq.~\eqref{eq:GM} for different temperatures below 1 GK indicates that the effective energy range for this reaction falls below 1 MeV. The recommended Maxwellian-averaged reaction rate $N_A\langle \sigma v \rangle$ for $pd$ radiative capture reaction at temperatures below 1 GK is presented in Table~\ref{tab:rate}. In comparison, Fig.~\ref{fig:rate} illustrates the ratio of rates obtained from Refs.~\cite{mossa2020,moscoso2021,descouvemont2004,xu2013,iliadis2016,coc2015} to the rates calculated in the present work. The dashed and solid curves in Fig.~\ref{fig:rate} represent the resulting reaction rates for the calculation without and with the inclusion of $M1$ transition, respectively. Notably, there is a significant difference between these two curves. Our calculation shows a good agreement with Refs.~\cite{iliadis2016,coc2015} at very low temperatures. Values from Refs.~\cite{iliadis2016,coc2015} exhibit no significant difference, as both adopted fitting based on theoretical $S$ factors \cite{marcucci2016}. The calculated rates are approximately $1.5\%$ higher than the NACRE II compilation \cite{xu2013} but the difference becomes larger at high temperatures. The $R$-matrix analysis \cite{descouvemont2004} and the recent best-fit measured $S$ factors \cite{mossa2020,moscoso2021} provide rates higher than our calculation below $0.1$ GK. The discrepancy between our calculated rates and those from Refs.~\cite{mossa2020,moscoso2021,iliadis2016,coc2015} is below $5\%$. 

\begin{table}
    \centering
    \caption{Calculated reaction rates in this work.}
    \label{tab:rate}
    \begin{tabular}{cccc}\\
    \hline \hline
    $T$ & $N_A\langle \sigma v \rangle$ & $T$ & $N_A\langle \sigma v \rangle$ \\
    (GK) & (cm$^{3}$ mol$^{-1}$ s$^{-1}$) & (GK) & (cm$^{3}$ mol$^{-1}$ s$^{-1}$) \\
    \hline
$0.001$	&	$9.445 \times 10^{-12}$	&	$0.002$	&	$1.877 \times 10^{-08}$	\\
$0.003$	&	$6.159 \times 10^{-07}$	&	$0.004$	&	$5.460 \times 10^{-06}$	\\
$0.005$	&	$2.558 \times 10^{-05}$	&	$0.006$	&	$8.271 \times 10^{-05}$	\\
$0.007$	&	$2.105 \times 10^{-04}$	&	$0.008$	&	$4.539 \times 10^{-04}$	\\
$0.009$	&	$8.677 \times 10^{-04}$	&	$0.010$	&	$1.514 \times 10^{-03}$	\\
$0.011$	&	$2.462 \times 10^{-03}$	&	$0.012$	&	$3.783 \times 10^{-03}$	\\
$0.013$	&	$5.551 \times 10^{-03}$	&	$0.014$	&	$7.840 \times 10^{-03}$	\\
$0.015$	&	$1.073 \times 10^{-02}$	&	$0.016$	&	$1.428 \times 10^{-02}$	\\
$0.018$	&	$2.368 \times 10^{-02}$	&	$0.020$	&	$3.658 \times 10^{-02}$	\\
$0.025$	&	$8.726 \times 10^{-02}$	&	$0.030$	&	$1.691 \times 10^{-01}$	\\
$0.040$	&	$4.426 \times 10^{-01}$	&	$0.050$	&	$8.774 \times 10^{-01}$	\\
$0.060$	&	$1.480 \times 10^{+00}$	&	$0.070$	&	$2.248 \times 10^{+00}$	\\
$0.080$	&	$3.179 \times 10^{+00}$	&	$0.090$	&	$4.264 \times 10^{+00}$	\\
$0.100$	&	$5.497 \times 10^{+00}$	&	$0.110$	&	$6.871 \times 10^{+00}$	\\
$0.120$	&	$8.379 \times 10^{+00}$	&	$0.130$	&	$1.001 \times 10^{+01}$	\\
$0.140$	&	$1.177 \times 10^{+01}$	&	$0.150$	&	$1.363 \times 10^{+01}$	\\
$0.160$	&	$1.561 \times 10^{+01}$	&	$0.180$	&	$1.987 \times 10^{+01}$	\\
$0.200$	&	$2.451 \times 10^{+01}$	&	$0.250$	&	$3.755 \times 10^{+01}$	\\
$0.300$	&	$5.236 \times 10^{+01}$	&	$0.350$	&	$6.865 \times 10^{+01}$	\\
$0.400$	&	$8.621 \times 10^{+01}$	&	$0.450$	&	$1.048 \times 10^{+02}$	\\
$0.500$	&	$1.245 \times 10^{+02}$	&	$0.600$	&	$1.662 \times 10^{+02}$	\\
$0.700$	&	$2.106 \times 10^{+02}$	&	$0.800$	&	$2.573 \times 10^{+02}$	\\
$0.900$	&	$3.059 \times 10^{+02}$	&	$1.000$	&	$3.561 \times 10^{+02}$	\\
    \hline \hline 
\end{tabular}
\end{table}

\section{Conclusions}
Our investigation has provided a quantitative analysis of the role of $M1$ transition in the $pd$ radiative capture process at extremely low energies, employing the potential model. The form of the phenomenological potential is derived from microscopic calculations, showcasing the effectiveness of the potential model as a simple yet powerful tool for addressing few-body problems. We have emphasized that the contribution of the $M1$ transition is highly sensitive to scattering potentials. Our calculated astrophysical $S$ factors, reaction rates, and elastic scattering observables closely align with recent works, showing a difference of less than 10\%. This good agreement reinforces the validity of our approach.

\ack

The authors would like to thank Bui Minh Loc for engaging in discussions.
This work was funded by Ho Chi Minh City University of Education Foundation for Science and Technology under grant number CS.2023.19.57. N. L. A. acknowledges the Master, PhD Scholarship Programme of Vingroup Innovation Foundation (VINIF), code VINIF.2023.TS.003.

\section*{References}

\bibliography{refs}

\end{document}